\documentclass[]{interact}

\usepackage{epstopdf}
\usepackage[caption=false]{subfig}

\usepackage[numbers,sort&compress]{natbib}
\bibpunct[, ]{[}{]}{,}{n}{,}{,}

\theoremstyle{plain}
\newtheorem{theorem}{Theorem}[section]
\newtheorem{lemma}[theorem]{Lemma}
\newtheorem{corollary}[theorem]{Corollary}

\theoremstyle{definition}
\newtheorem{definition}[theorem]{Definition}

\theoremstyle{remark}
\newtheorem{remark}{Remark}

\def\bra#1{\langle#1|}
\def\braket#1#2{\langle#1|#2\rangle}
\def\ket#1{|#1\rangle}

\def\C{\mathbb{C}}
\def\F{\mathbb{F}}
\def\Pr{\mathop{\rm Pr}\nolimits}
\def\tr{\mathop{\rm tr}\nolimits}
\def\Tr{\mathop{\rm Tr}\nolimits}
\def\wgt{\mathop{\rm wgt}\nolimits}
\def\linspan{\mathop{\rm span}\nolimits}

\begin{document}


\title{Algebraic Quantum Codes:\\
Linking Quantum Mechanics and Discrete Mathematics}

\author{
\name{Markus Grassl\thanks{Email: markus.grassl@ug.edu.pl}}
\affil{International Centre for Theory of Quantum
  Technologies, University Gdansk, Gda\'nsk, Poland}
}

\def\today{November 13, 2020}

\maketitle

\begin{abstract}
We present a general framework of quantum error-correcting codes
(QECCs) as a subspace of a complex Hilbert space and the corresponding
error models. Then we illustrate how QECCs can be constructed using
techniques from algebraic coding theory. Additionally, we discuss
secondary constructions for QECCs, leading to propagation rules for
the parameters of QECCs.
\end{abstract}


\section{Introduction}
Quantum error-correcting codes (QECCs) are essential for the
realization of information processing using the principles of quantum
mechanics. About 25 years ago, Peter Shor presented the first scheme
to reduce errors in a quantum computer \cite{Sho95}. This was followed
by more general constructions by Calderbank and Shor \cite{CaSh96} as
well as Steane \cite{Ste96:error}, now referred to as CSS codes.

The ground for the theory of so-called stabilizer codes was laid by
the work of Gottesman \cite{Got96_hamming} from the physics point of
view, and by Calderbank, Rains, Shor, and Sloane \cite{CRSS98} taking
more the point of view of algebraic codes.  In a nutshell, the theory
of stabilizer codes allows the construction of quantum
error-correcting codes using classical codes that are self-orthogonal
with respect to a certain symplectic inner product.  

A survey on constructions of stabilizer codes from various families of
classical codes can be found in \cite{KKKS06}.  A collection of
quantum codes with the best known parameters is available online
\cite{Grassl:codetables}. Currently, the table covers only so-called
qubit codes, but it is planned to extend them in the near future.

In this article, we focus on the basic principles of general quantum
error-correcting codes and establish the link between quantum
mechanics and algebraic coding theory.  Additionally, we discuss
secondary constructions of quantum codes, i.\,e., how one can derive new
quantum codes from a given one.

\section{General Quantum Error-Correcting Codes}
\subsection{Axiomatic Quantum Mechanics}
In order to establish the framework for quantum error-correcting
codes, we introduce the required concepts of quantum mechanics in an
axiomatic way.  A more detailed description can, for example, be found
in the book by Nielsen and Chuang \cite{NiCh00}.

\subsubsection{Pure quantum states}
To every quantum mechanical system, we associate a complex Hilbert
space $\mathcal{H}$. In our context, the dimension of the Hilbert
space is finite, i.\,e., $\mathcal{H}=\C^d$ for some positive integer
$d$. The basis states of the Hilbert space correspond to perfectly
distinguishable states of the quantum mechanical system. For the
smallest example $d=2$, the two basis states may correspond to an atom
or ion being in its ground or exited state, a quantum mechanical spin
being aligned parallel or anti-parallel with an external magnetic
field, or two orthogonal polarization directions of a photon. When we
have complete knowledge of the state of the quantum mechanical system,
the system is in a \emph{pure state} that can be described by a
unit-norm vector in the Hilbert space $\mathcal{H}$.  Looking again at
the case $d=2$, the state of a so-called \emph{qubit} is given by
\begin{alignat}{5}\label{eq:qubit}
\ket{\psi} = \alpha\ket{0}+\beta\ket{1},\qquad|\alpha|^2+|\beta|^2=1
\end{alignat}
with complex coefficients $\alpha,\beta\in\C$. Here we have used the
\emph{ket-notation} for the column vectors of an orthonormal basis of
$\mathcal{H}$, i.\,e., 
\begin{alignat}{5}
\ket{0}\hat{=}\begin{pmatrix}1\\0\end{pmatrix}
\qquad\text{and}\qquad
\ket{1}\hat{=}\begin{pmatrix}0\\1\end{pmatrix}.
\end{alignat}
When both coefficients in \eqref{eq:qubit} are non-zero, $\alpha\ne 0
\ne\beta$, the state is referred to as a \emph{superposition} of the
basis states $\ket{0}$ and $\ket{1}$.

Correspondingly, the state of a so-called \emph{qudit}, a
$d$-dimensional quantum system, is given by
\begin{alignat}{5}\label{eq:qudit}
\ket{\psi} = \sum_{i=0}^{d-1} c_i\ket{i},
  \qquad\text{with $\sum_{i=0}^{d-1}|c_i|^2=1$},
\end{alignat}
where the states $\{\ket{0},\ket{1},\ldots,\ket{d-1}\}$ form an
orthonormal basis of $\mathcal{H}=\C^d$.  When the dimension $d=p^m=q$
is a prime power, we can label the basis states by elements of the
finite field $\F_q$ instead of integers:
\begin{alignat}{5}\label{eq:qudit_Fq}
\ket{\psi} = \sum_{x\in\F_q} c_x\ket{x},
  \qquad\text{with $\sum_{x\in\F_q}|c_x|^2=1$.}
\end{alignat}
One may think of the state $\ket{\psi}$ as an element of the group
algebra $\C[\F_q]$.

The state space of composite quantum systems is given by the tensor
product of the state spaces of the individual quantum systems, i.\,e.,
for a bipartite quantum system, we have
$\mathcal{H}_{12}=\mathcal{H}_1\otimes \mathcal{H}_2$. When a system
is composed of $n$ qudits, each of dimension $q$, the dimension of the
composite system is $q^n$, i.\,e., it grows exponentially in the number
of systems. A basis for the composite systems is given by the tensor
product of the bases for the component systems. The basis vectors
\begin{alignat}{5}
\ket{\bm{x}}
=\ket{x_1 x_2\ldots x_n}
=\ket{x_1}\otimes\ket{x_2}\otimes\dots\otimes\ket{x_n}
\end{alignat}
can be labeled by strings $\bm{x}=(x_1,x_2,\ldots,x_n)\in\F_q^n$ of
length $n$ over the finite field. The state of a \emph{quantum
  register} with $n$ qudits is given by
\begin{alignat}{5}\label{eq:qregister}
\ket{\psi} = \sum_{\bm{x}\in\F_q^n} c_{\bm{x}}\ket{\bm{x}},
  \qquad\text{with $\sum_{\bm{x}\in\F_q^n}|c_{\bm{x}}|^2=1$.}
\end{alignat}
Again, one may think of the state $\ket{\psi}$ being an element of the group
algebra $\C[\F_q^n]$.

\subsubsection{Quantum measurements and quantum operations}
From the physics point of view, a quantum measurement is associated
with a Hermitian (self-adjoint) operator $A$.  Being self-adjoint, the
eigenvalues $\lambda_i$ of the \emph{observable} $A$ are real and
correspond to some physical quantity, for example energy. As already
mentioned, the two basis states of a qubit may correspond to two
distinct energy states of an atom or ion.  More abstractly, any
Hermitian operator $A$ on a Hilbert space of dimension $d$ possesses a
\emph{spectral decomposition} of the form
\begin{alignat}{5}\label{eq:spectral}
 A =\sum_{i=0}^{d-1} \lambda_i \ket{\psi_i}\bra{\psi_i},
\end{alignat}
where the eigenvalues are $\lambda_i$, and $\ket{\psi_i}\bra{\psi_i}$
denotes the orthogonal projection onto the eigenspace spanned by the
corresponding eigenvector $\ket{\psi_i}$. Here $\bra{\psi_i}$ denotes
a \emph{bra-vector} that is the row vector formed by the complex
conjugate coefficients of the column vector $\ket{\psi_i}$. When the
spectrum of $A$ is degenerate, i.\,e., the eigenvalues are not all
distinct, the spectral decomposition \eqref{eq:spectral} can be
rewritten as
\begin{alignat}{5}\label{eq:spectral2}
 A =\sum_{i=0}^{m-1} \lambda_i P_i=
\sum_{i=0}^{m-1} \lambda_i \sum_{j=1}^{\mu_i}\ket{\psi_i^{(j)}}\bra{\psi_i^{(j)}}.
\end{alignat}
Here we assume that there are $m$ distinct eigenvalues $\lambda_i$
with multiplicities $\mu_i$. The vectors $\ket{\psi_i^{(j)}}$ form an
orthonormal basis of the corresponding eigenspace.

According to the postulates of quantum mechanics, when performing a
measurement of the observable $A$ on a quantum system in state
$\ket{\psi}$, one will observe a randomly chosen eigenvalue
$\lambda_i$. The probability to observe $\lambda=\lambda_i$ is given
by
\begin{alignat}{5}\label{eq:measurement_prob}
\Pr_{\ket{\psi}}(\lambda=\lambda_i) = \| P_i \ket{\psi}\|^2=\bra{\psi}P_i\ket{\psi}.
\end{alignat}
The post-measurement state is given by the re-normalized projection of
the state onto the corresponding eigenspace, i.\,e.,
\begin{alignat}{5}\label{eq:postmeasurement}
\ket{\psi(\lambda=\lambda_i)} = \frac{P_i\ket{\psi}}{\|P_i\ket{\psi}\|}.
\end{alignat}
In the context of quantum information processing, we are usually not
interested in the physical quantity corresponding to the eigenvalues
$\lambda_i$, but only in the index $i$ of the eigenvalue and the
associate eigenspace, given by the projection $P_i$.  We may, for
example, talk about \emph{measuring a quantum system in the standard
basis} when the eigenspaces are the one-dimensional spaces defined by
the basis of the Hilbert space.  For a qubit in the state
$\ket{\psi}=\alpha\ket{0}+\beta\ket{1}$ (cf. \eqref{eq:qubit}), we
will obtain the outcomes ``$0$'' and ``$1$'' corresponding to the
basis states $\ket{0}$ and $\ket{1}$, respectively, with probability
$|\alpha|^2$ and $|\beta^2|$, respectively. The fact that the
probabilities have to sum to unity explains why the vector describing
a pure quantum state has to be normalized.  In principle, one can also
start with a decomposition of the whole Hilbert space into mutually
orthogonal subspaces and define an observable that has the
corresponding eigenspaces.

Transformations of the state of a closed quantum system are described
by linear operators. As the operations have to preserve normalization,
the admissible transformations are unitary transformations $U\in {\rm
  U}(d)$. Note that the probability \eqref{eq:measurement_prob} of
observing a particular eigenvalue $\lambda_i$ of an observable $A$
does not change when multiplying the state vector by complex number
$e^{i\phi}$ of modulus $1$. Therefore, states that differ by a
\emph{global phase factor} $e^{i\phi}$ can not be
distinguished. Accordingly, the effective group of transformations of
quantum states are all elements of the special unitary group ${\rm
  SU}(d)$.

For composite quantum systems, operations that act on the individual
subsystems are termed \emph{local operations}. For an $n$-qudit system,
the group of local unitary transformations is given by the $n$-fold
tensor product ${\rm SU}(d)^{\otimes n}={\rm SU}(d)\otimes\ldots\otimes{\rm SU}(d)$.

Similarly, one can perform local measurements on the individual
subsystems. Given an observable $A$ on a single qudit, the operator
$A^{(1)} = A \otimes I_d\otimes\ldots\otimes I_d$ (where $I_d$ denotes
a $d\times d$ identity matrix) is the Hermitian operator corresponding
to the measurement of only the first qudit.  As an example, consider
the state of a two-qubit system given by
\begin{alignat}{5}
\ket{\psi}
&{}=c_{00}\ket{00}+c_{01}\ket{01}+c_{10}\ket{10}+c_{11}\ket{11}\\
&{}=\alpha\ket{0}\otimes\left(c'_{00}\ket{0}+c'_{01}\ket{1}\right)
    +\beta\ket{1}\otimes\left(c'_{10}\ket{0}+c'_{11}\ket{1}\right).
\end{alignat}
The coefficients in the second form are chosen such that the states of
the second qubit are normalized, i.\,e.,
$|c'_{00}|^2+|c'_{01}|^2=|c'_{10}|^2+|c'_{11}|^2=1$,which implies $|\alpha|^2+|\beta|^2=1$.
Hence the probability to obtain the result ``$0$'' or ``$1$'' when
measuring the first qubit is given by $|\alpha|^2$ and $|\beta|^2$,
respectively. The corresponding post-measurements states are
\begin{alignat}{5}
\ket{\psi(y=0)}&{}=\ket{0}\otimes\left(c'_{00}\ket{0}+c'_{01}\ket{1}\right)\\
\text{and}\quad
\ket{\psi(y=1)}&{}=\ket{1}\otimes\left(c'_{10}\ket{0}+c'_{11}\ket{1}\right).
\end{alignat}
For the \emph{maximally entangled} state
\begin{alignat}{5}
\ket{\Psi^+}=\frac{1}{\sqrt{2}}\ket{00}+\frac{1}{\sqrt{2}}\ket{11},
\end{alignat}
the probability for each outcome is $1/2$, regardless which of the
qubits is measured in the standard basis. The possible
post-measurement states are $\ket{00}$ and $\ket{11}$. Hence,
measuring the second qubit after the first qubit has been measured,
the result will always agree with the outcome of the first
measurement. We will not discuss this phenomenon of
\emph{entanglement} in more detail here, but it has to be noted that
measuring one subsystem affects the state of the other subsystem.

\subsubsection{Mixed quantum states and reduced quantum states}
When performing a measurement on quantum systems, each of the
post-measurement states occurs with some probability.  When we ignore
the measurement result, we have an ensemble of quantum states
$\{\ket{\psi_i}\}$ with corresponding probabilities $p_i$. Such an
ensemble can be described by the \emph{density matrix}
\begin{alignat}{5}\label{eq:densitymatrix}
\rho=\sum_i p_i\ket{\psi_i}\bra{\psi_i},
\end{alignat}
where $\ket{\psi_i}\bra{\psi_i}$ is the orthogonal projection onto the
state $\ket{\psi_i}$. This concept holds more generally, e.\,g., when we
prepare different quantum states $\ket{\psi_i}$ with some probability
$p_i$ or when a source emits quantum states with certain
probabilities.  Different ensembles of quantum states may result in
the same density matrix $\rho$, but the density matrix contains all
information that can be obtained from the ensemble.  When we start
with a uniform distribution over the states of a basis of the Hilbert
space, the corresponding density matrix will be proportional to the
identity matrix:
\begin{alignat}{5}
\rho = \frac{1}{d} I.
\end{alignat}
This \emph{maximally mixed} state is invariant under unitary
transformations, i.\,e., it is uniformly random with respect to any
basis.  Another special case is an ensemble with just a single pure
state $\ket{\psi}$. The corresponding density matrix equals the
projection onto the one-dimensional space spanned by the state
$\ket{\psi}$, i.\,e., 
\begin{alignat}{5}
\rho = \ket{\psi}\bra{\psi}.
\end{alignat}

As we have seen, measuring a subsystem of a composite quantum system
has also an effect on the subsystems that we have not measured. If we
ignore the measured subsystem and the measurement result, we arrive at
an ensemble of quantum states for the remaining subsystems, described
by the \emph{reduced density matrix}. To make this more precise,
consider a bipartite quantum system of dimension $d_1 d_2$. A mixed
state of the composite system can be written as
\begin{alignat}{5}\label{eq:rho12}
\rho_{12}=\sum_{i,i'=0}^{d_1-1}\sum_{j,j'=0}^{d_2-1} c_{i,j,i',j'}\ket{i,j}\bra{i',j'}.
\end{alignat}
Measuring the first system in the standard basis and discarding both
the measurement result and the first system, the \emph{reduced quantum state}
of the second system is given by 
\begin{alignat}{5}\label{eq:reducedrho}
\rho_{2}=\Tr_1(\rho)=\sum_{j,j'=0}^{d_2-1} \sum_{i=0}^{d_1-1} c_{i,j,i,j'}\ket{j}\bra{j'}.
\end{alignat}
The matrix $\rho_2$ is referred to as the \emph{partial trace} of
$\rho$ with respect to system $1$, or \emph{reduced density matrix} of
system $2$.  It can be obtained by summing the $d_1$ non-overlapping
submatrices of size $d_2\times d_2$ on the diagonal of $\rho$, when we
interpret the first index $i$ as labeling the block and the second
index $j$ as labeling the position within the block.  Note that from
the state of the second subsystem only, one cannot deduce whether a
measurement has been performed on the first system or not. The reduced
density matrix \eqref{eq:reducedrho} is independent of that
measurement.

While the partial trace discards all information of the first system,
one can define an inverse of the operation \eqref{eq:reducedrho} in
the following sense. For any mixed state $\rho$ on a $d$-dimensional
Hilbert space $\mathcal{H}_1$, there exists a pure state
$\ket{\psi}_{12}$ on the joint Hilbert space
$\mathcal{H}_1\otimes\mathcal{H}_2$ with $\dim\mathcal{H}_2=d$ such
that
\begin{alignat}{5}\label{eq:purification}
\rho_1 = \Tr_2(\ket{\psi}\bra{\psi}).
\end{alignat}
This \emph{purification} $\ket{\psi}_{12}$ of the mixed state $\rho_1$
is unique up to a unitary transformation on the Hilbert space $\mathcal{H}_2$.

\subsection{Characterizing Quantum Codes}
\subsubsection{Quantum channels}
The notion of error correction requires to fix an error model. In
quantum information processing, the two main sources of errors are
limited precision of the desired operations and unwanted interaction
of the system with its environment. Imperfect operations can be
modeled as operations that depend on the state of the environment.

As described above, using the purification of mixed states we can
always choose the Hilbert space of the system of sufficiently large
dimension so that the state of the system $\ket{\psi}_{\text{sys}}$ is
pure. Similarly, we can assume that the environment is initially in
the pure state $\ket{\varepsilon}_{\text{env}}$. Additionally, we make
the assumption that system and environment have not yet interacted,
and hence the initial state is described by the tensor product
\begin{alignat}{5}
\ket{\Psi}_{\text{in}}=\ket{\psi}_{\text{sys}}\otimes\ket{\varepsilon}_{\text{env}}.
\end{alignat}
The interaction of system and environment is given by a unitary
transformation $U_{\text{sys/env}}$  on the joint Hilbert space,
resulting in the state
\begin{alignat}{5}
\ket{\Psi}_{\text{out}}=U_{\text{sys/env}}\bigl(\ket{\psi}_{\text{sys}}\otimes\ket{\varepsilon}_{\text{env}}\bigr).
\end{alignat}
As we have no access to the environment, we take the partial trace
over the environment and obtain the reduced density matrix of the
system
\begin{alignat}{5}\label{eq:unitary_channel}
\rho_{\text{out}}=\Tr_{\text{env}}
\left(U_{\text{sys/env}}\left(\ket{\psi}\bra{\psi}_{\text{sys}}
\otimes\ket{\varepsilon}\bra{\varepsilon}_{\text{env}}\right)U_{\text{sys/env}}^{-1}\right).
\end{alignat}
Fixing the initial state $\ket{\varepsilon}_{\text{env}}$ of the
environment and the unitary interaction $U_{\text{sys/env}}$, one can
rewrite \eqref{eq:unitary_channel} in the form
\begin{alignat}{5}
\rho_{\text{out}}=\sum_i E_i \ket{\psi}\bra{\psi}_{\text{sys}} E_i^\dagger,
\end{alignat}
where $E_i^\dagger$ denotes the adjoint (complex conjugate transpose) of
the matrix $E_i$. Clearly, the matrices $E_i$ depend on
$\ket{\varepsilon}_{\text{env}}$ and $U_{\text{sys/env}}$.

\begin{definition}[quantum channel]
A \emph{quantum channel} $\mathcal{Q}$ is a linear transformation on
mixed quantum states of the form
\begin{alignat}{5}
\rho\mapsto \mathcal{Q}(\rho) = \sum_i E_i \rho E_i^\dagger.
\end{alignat}
The operators $E_i$ are referred to as \emph{error operators} or
\emph{Kraus operators} of the channel.
\end{definition}
The analog of the binary symmetric channel---or more generally, the
uniform symmetric channel or the additive white Gaussian noise (AWGN)
channel---is the \emph{depolarizing channel} given by
\begin{alignat}{5}\label{eq:depolarizing}
\mathcal{Q}_{\text{depol}}(\rho) = (1-p)\rho+p \frac{1}{d}I.
\end{alignat}
With probability $p$, the depolarizing channel replaces the input
state by the maximally mixed state, and with probability $1-p$, the
input state is not changed. Like its classical counterpart, the
depolarizing channel is in some sense related to the worst case
assumption that the channel either transmits the input error-free or
replaces it with a completely random state.

Another important example is the \emph{quantum erasure channel}
\cite{GBP97} for which the dimension of the Hilbert spaces for the
input and output differ. With probability $p$, the input is replaced
by a state $\ket{\bot}$ that is orthogonal to all states of the input
Hilbert space, and with probability $1-p$ the state is transmitted
without error:
\begin{alignat}{5}\label{eq:erasure}
\mathcal{Q}_{\text{erasure}}(\rho) = (1-p)\rho+p \ket{\bot}\bra{\bot}.
\end{alignat}
The state $\ket{\bot}$ indicates that the quantum information has been
erased.

The combination of two independent channels is modeled as follows:
\begin{definition}[product channel]\label{def:product_channel}
Given two quantum channels $\mathcal{Q}_1$ and $\mathcal{Q}_2$ with
error operators $E_i$ and $F_j$, respectively, the action of the
\emph{product channel} $\mathcal{Q}_1\otimes \mathcal{Q}_2$ on a mixed
stated $\rho_{12}$ of the composite system (cf. \eqref{eq:rho12}) is
given by
\begin{alignat}{5}
\rho_{12}\mapsto\left(\mathcal{Q}_{2}\otimes\mathcal{Q}_2\right)(\rho_{12})
   = \sum_{i,j} (E_i\otimes F_j) \rho_{12} (E_i^\dagger\otimes F_j^\dagger).
\end{alignat}
\end{definition}
A particular example is the channel $\mathcal{Q}^{\otimes n}$ which
models $n$ independent uses of the memoryless channel $\mathcal{Q}$.
Loosely speaking, this requires that the Hilbert space of the
environment is different for each use of the channel.  
\subsubsection{Necessary and sufficient conditions for quantum error correction}
While a classical error-correcting code uses only a subset of all
possible messages, a quantum error-correcting code (QECC) uses only
states in a suitably chosen subspace of the full Hilbert space. Given
a quantum channel $\mathcal{Q}$ with error operators $E_i$, we have
the following necessary and sufficient conditions for perfect error
correction, also known as the \emph{Knill-Laflamme conditions}
\cite{KnLa97}:
\begin{theorem}\label{thm:KnillLaflamme}
A subspace $\mathcal{C}$ with orthonormal basis $\{ \ket{c_i}\}$ of a
Hilbert space $\mathcal{H}$ is a quantum error-correcting code for a
quantum channel with error operators $\{ E_k\}$ if and only if the
following conditions hold for all $i,j,k,\ell$:
\begin{alignat}{6}
&\text{\rm(i)} \qquad& \bra{c_i} E_k^\dagger E_\ell \ket{c_j} &{}=
  0&\text{for $i\ne j$}\label{eq:KnillLaflamme1}\\
&\text{\rm(ii)} & \bra{c_i} E_k^\dagger E_\ell \ket{c_i} &{}= \bra{c_j} E_k^\dagger E_\ell \ket{c_j}=\alpha_{k\ell}\label{eq:KnillLaflamme2}
\end{alignat}
In terms of the orthogonal projection
$P_{\mathcal{C}}=\sum_i\ket{c_i}\bra{c_i}$ onto the code
$\mathcal{C}$, the conditions are equivalent to
\begin{alignat}{6}
&\text{\rm(iii)} \qquad& P_{\mathcal{C}} E_k^\dagger E_\ell P_{\mathcal{C}}=\alpha_{k\ell}P_{\mathcal{C}}.\label{eq:KnillLaflamme3}
\end{alignat}
\end{theorem}
The first condition \eqref{eq:KnillLaflamme1} requires that orthogonal
states in the code---which can be perfectly distinguished by a
suitable measurement---remain orthogonal under the action of the
channel. Loosely speaking, the second condition
\eqref{eq:KnillLaflamme2} requires that errors transform all states of
the code in the same way.

First we note without proof that the Knill-Laflamme conditions are bi-linear in the
error operators $E_k$:
\begin{lemma}
If the conditions \eqref{eq:KnillLaflamme1} and \eqref{eq:KnillLaflamme2}
hold for error operators $E_k$ and $E_\ell$, then they hold for any
linear combination $\mu E_k+\nu E_\ell$.
\end{lemma}
This implies the following.
\begin{corollary}\label{cor:linearity}
It is sufficient that the conditions \eqref{eq:KnillLaflamme2} and
\eqref{eq:KnillLaflamme1} hold for a basis of the linear space of
operators spanned by the error operators $\{ E_k\}$ of the quantum
channel.
\end{corollary}
Since we are considering only finite dimensional Hilbert spaces here,
it turns out that we are dealing with a finite number of errors that
have to be corrected, while the error operators of the channel might
be parameterized by continuous parameters. It can be shown that one can
replace the original error operators by linear combinations such that
the Knill-Laflamme conditions have a particularly nice form (see,
e.\,g., \cite{Gra02}):
\begin{theorem}
Let $\mathcal{C}$ with orthonormal basis $\{\ket{c_i}\}$ be a QECC for
the channel $\mathcal{Q}$ with errors operators $\{ E_k\}$. Then there
are error operators $\{\widetilde{E}_k\}$ such that
\begin{alignat}{5}
\bra{c_i}\widetilde{E}_k^\dagger\widetilde{E}_\ell\ket{c_j}=\delta_{ij}\delta_{k\ell},\label{eq:QECC_decomp}
\end{alignat}
and the linear span of $\{E_k\ket{c_i}\}$ equals the linear span of $\{\widetilde{E}_k\ket{c_i}\}$.
\end{theorem}
Using \eqref{eq:QECC_decomp}, it turns out that the images
\begin{alignat}{5}
 \mathcal{V}_k = \widetilde{E}_k\mathcal{C}=\linspan\{ \widetilde{E}_k\ket{c_i}\}
\end{alignat}
of the code $\mathcal{C}$ under the error operators $\widetilde{E}_k$
are mutually orthogonal. Moreover, for fixed $k$ the states
$\{\widetilde{E}_k\ket{c_i}\}$ constitute an orthonormal basis of
$\mathcal{V}_k$, and all those spaces have the same dimension as the
code $\mathcal{C}$. Hence there are isometries mapping the space
$\mathcal{V}_k$ to the code $\mathcal{C}$. There is a measurement
whose eigenspaces are the spaces $\mathcal{V}_k$. Performing that
measurement yields information onto which of the spaces $\mathcal{V}_k$
the output of the quantum channel has been projected.  Applying the
corresponding isometry, the state can be mapped back to the code
$\mathcal{C}$, i.\,e., perfectly corrected.

\subsubsection{Local errors and error bases}
For classical error correcting-codes, the Hamming weight of an error
equals the number of positions that are changed by the error. We can
define a similar notion for quantum codes $\mathcal{C}$ for the
$n$-fold use of a quantum channel $\mathcal{Q}$ acting on a Hilbert
space $\mathcal{H}=\C^q$. In this case, the error operators of the
product channel $\mathcal{Q}^{\otimes n}$ are tensor products of the
error operators of the channel $\mathcal{Q}$ (see Definition
\ref{def:product_channel}). We distinguish whether an error acts
trivially on a certain subsystem, i.\,e., is proportional to identity, or
not. We extend this notion by linearity and obtain:
\begin{definition}[error weight]\label{def:errorweight}
The \emph{weight} $\wgt(E)$ of an error operator $E$ acting on the
Hilbert space $\left(\C^{q}\right)^{\otimes
  n}=\C^q\otimes\ldots\otimes \C^q$ equals the number of subsystems on
which it acts non-trivially.
\end{definition}
If $E = I_{q^{n-t}}\otimes E'$, where $I_{q^{n-t}}$ is an identify
matrix of size $q^{n-t}\times q^{n-t}$, the weight of $E$ is at most
$t$, the number of subsystems on which $E'$ acts. In particular, when
$E=E_1\otimes E_2\otimes\ldots\otimes E_n$, the weight of $E$ equals
the number of tensor factors that are not proportional to identity.

\begin{definition}
A quantum error-correcting code $\mathcal{C}=(\!(n,K,d)\!)_q$ is a
subspace of dimension\footnote{For simplicity, we exclude the case $K=1$
  here.} $K>1$ of the $n$-fold tensor product
$\left(\C^q\right)^{\otimes n}$ such that the Knill-Laflamme
conditions \eqref{eq:KnillLaflamme1} and \eqref{eq:KnillLaflamme2}
hold for all pairs of errors $E_k$, $E_\ell$ with $\wgt(E_k^\dagger
E_\ell)<d$.
\end{definition}
When the true \emph{minimum distance} of the code $\mathcal{C}$ equals
$d$, then there is an operator $E=E_k^\dagger E_\ell$ of weight $d$
such that for two orthonormal states $\ket{c_i}$ and $\ket{c_j}$ of
the code, the image of $\ket{c_j}$ under $E$ overlaps with
$\ket{c_i}$, i.\,e., $\bra{c_i}E\ket{c_j}\ne 0$, or the action of $E$
relative to $\ket{c_i}$ and $\ket{c_j}$ differs, i.\,e.,
$\bra{c_i}E\ket{c_i}\ne \bra{c_j}E\ket{c_j}$. The latter situation
occurs, for example, when both states are eigenstates of $E$, but with
different eigenvalues.  On the other hand, there might be errors $E$
of weight $0<\wgt(E)<d$ that act trivially on the code, i.\,e., the
restriction of $E$ to the code subspace is proportional to identity
(see \eqref{eq:KnillLaflamme3}). Clearly, such an ``error'' $E$,
having no effect on the code, can not be detected.

When we only distinguish whether an error operator $E$ on a subsystem
of dimension $q$ is proportional to identity or not, the possible
errors are all linear operators. The dimension of the vector space of
those operators is $q^2$. According to Corollary \ref{cor:linearity},
it is sufficient to consider a basis of $q^2$ operators.  Naturally,
the basis is chosen to contain identity.  Recall that when the
dimension $q$ of the subsystem is a prime power $q=p^m$, we can label
the basis states by elements of the finite field $\F_q$.
\begin{lemma}[local error basis]\label{lemma:errorbasis}
On the Hilbert space $\C^q$ with orthonormal basis $\{\ket{x}\colon
  x\in\F_q\}$ we define the operators
\begin{alignat}{5}
X^a &{}= \sum_{x\in\F_q}\ket{x+a}\bra{x},\\
Z^b &{}= \sum_{y\in\F_q}\omega_p^{\tr(by)}\ket{y}\bra{y},
\end{alignat}
where $\omega_p=\exp(2\pi i/p)$ is a complex primitive $p$-th root of
unity and $\tr(y)$ denotes the absolute trace from $\F_q$ to the prime
field $\F_p$.
Then the  $q^2$ operators $\{X^a Z^b\colon a,b\in\F_q\}$ form a basis
of all linear operators on $\C^q$, containing the operator $I=X^0Z^0$.
\end{lemma}
\begin{proof}
In order to show that the operators form a basis, one checks that they
are orthogonal with respect to $\Tr(A^\dagger B)$.
\end{proof}
The operators $X^a$ correspond to permutations of the basis states
according to the addition of $a$. The operators $Z^b$ are their
diagonalized version, and the operators $X^a$ and $Z^b$ are related by
the Fourier transformation for the additive group of $\F_q$.
Taking tensor products of the elements of the local error basis with
no more than $t$ operators different from identity, one obtains a
basis of all errors up to weight $t$.  Generalizations of error bases
with nice properties can be found in \cite{KlRo02:UEB1}.

In summary, Theorem \ref{thm:KnillLaflamme} characterizes quantum
error-correcting codes for arbitrary channels. Considering $n$
independent uses of the same channel, one can introduce the notion of
the weight of an error and arrives at the more special case of a
quantum error-correcting code $\mathcal{C}=(\!(n,K,d)\!)_q$ being a
linear subspace of $\left(\C^q\right)^{\otimes n}$ with
$\dim\mathcal{C}=K$. Such a code is able to
\begin{itemize}
\item correct all errors $E$ of weight $\wgt(E)< d/2$, or
\item correct up to  $t < d$ erasures, or
\item detect all errors $E$ of weight $\wgt(E) < d$ that act
  non-trivially on the code.
\end{itemize}
Operationally, the parameter $d$ plays the same role as the minimum
distance of classical codes. A discussion of the assumption of local
errors can be found, e.\,g., in \cite{KLZ98}.

\section{Stabilizer Codes}
The general theory of quantum error-correcting codes described in the
previous section does not provide information on how to construct good
codes. In this section we will establish the link between quantum and
classical codes.
\subsection{The Generalized $n$-qudit Pauli Group}
We return to the local error basis of Lemma \ref{lemma:errorbasis} and
notice that the operators $X^a$ and $Z^b$ generate a matrix group
$\mathcal{P}=\langle X^a Z^b\colon a,b\in\F_q\rangle$ which is
sometimes referred to as the \emph{generalized Pauli group}. The order
of the group is $|\mathcal{P}|=pq^2$, and the center consists of the
multiples of identity $Z(\mathcal{P})=\{\exp(k2\pi i/p)I\colon k
=0,\ldots,p-1\}$. Conjugating an element $Z^b$ by an element $X^a$
results in a phase factor:
\begin{alignat}{5}
X^{-a} Z^b X^a 
&{}=
\left( \sum_{z\in\F_q}\ket{z-a}\bra{z}\right)
\left( \sum_{y\in\F_q}\omega_p^{\tr(by)}\ket{y}\bra{y} \right)
\left( \sum_{x\in\F_q}\ket{x+a}\bra{x}\right)\\
&{}=\sum_{x,y,z\in\F_q}
\omega_p^{\tr(by)}\ket{z-a}\braket{z}{y}\braket{y}{x+a}\bra{x}\\
&{}=\sum_{x\in\F_q}
\omega_p^{\tr(b(x+a))}\ket{x}\bra{x}=\omega_p^{\tr(ab)} Z^b\label{eq:phase}
\end{alignat}
Here we have used the linearity of the trace and the fact that the
basis states are orthonormal, i.\,e.,
$\braket{x}{y}=\delta_{x,y}$. Eq. \eqref{eq:phase} implies the
relation
\begin{alignat}{5}
Z^b X^a = X^a\left(X^{-a} Z^b X^a\right) = \omega_p^{\tr(ab)} X^a Z^b,
\end{alignat}
and further
\begin{alignat}{5}
X^aZ^b X^{a'}Z^{b'} 
&{}= \omega_p^{\tr(a'b)} X^a X^{a'} Z^b Z^{b'}
= \omega_p^{\tr(a'b)} X^{a'}X^a Z^{b'} Z^b\\
&{}= \omega_p^{\tr(a'b-ab')} X^{a'} Z^{b'} X^a Z^b.\label{eq:commutator}
\end{alignat}
Forming $n$-fold tensor products of the local error operators, we get
the generalized $n$-qudit Pauli group $\mathcal{P}_n$ of order
$|\mathcal{P}_n|=pq^{2n}$. The elements of the group can be written as
\begin{alignat}{5}
\mathcal{P}_n =\{ \omega_p^\gamma X^{a_1}Z^{b_1}\otimes\ldots\otimes
X^{a_n}Z^{b_n}\colon a_i,b_i\in\F_q,\gamma=0,\ldots,p-1\}.
\end{alignat}
Modulo its center, the group $\mathcal{P}_n$ is isomorphic to the
vector space $\F_q^{2n}$ considered as an additive group via the
homomorphism
\begin{alignat}{5}
\phi\colon 
\omega_p^\gamma X^{a_1}Z^{b_1}\otimes\ldots\otimes X^{a_n}Z^{b_n}\mapsto
(a_1,\ldots a_n|b_1,\ldots,b_n)=(\bm{a}|\bm{b}).\label{eq:homomorph}
\end{alignat}
What is more, using the commutator relation \eqref{eq:commutator} for
single-qudit operators, we obtain the commutator relation
\begin{alignat}{5}
X^{\bm{a}}Z^{\bm{b}} X^{\bm{a}'}Z^{\bm{b}'} = \omega_p^{\tr(\bm{a}'\bm{b}-\bm{a}\bm{b}')}X^{\bm{a'}}Z^{\bm{b'}} X^{\bm{a}}Z^{\bm{b}}.
\label{eq:commutator_n}
\end{alignat}
Here we have used the shorthand
$X^{\bm{a}}=X^{a_1}\otimes\ldots\otimes X^{a_n}$ and
$Z^{\bm{b}}=Z^{b_1}\otimes\ldots\otimes Z^{b_n}$.  The exponent of the
phase factor in \eqref{eq:commutator_n} is determined by the
$\F_p$-valued symplectic form
\begin{alignat}{5}
(\bm{a}|\bm{b}) * (\bm{a}'|\bm{b}') = \tr(\bm{a}'\cdot\bm{b}-\bm{a}\cdot\bm{b}')
=\tr\left(\sum_{i=1}^na_i'b_i-a_ib_i'\right).\label{eq:symplectic}
\end{alignat}
on $\F_q^{2n}=\F_q^n\times\F_q^n$.

The homomorphism $\phi$ in \eqref{eq:homomorph} is compatible with the
weight of an operator in Definition \ref{def:errorweight}:
\begin{alignat}{5}
\wgt\left(X^{\bm{a}}Z^{\bm{b}}\right) = |\{ i\colon i=1,\ldots,n|
(a_i,b_i)\ne (0,0)\}|.\label{eq:symplecticweight}
\end{alignat}
The right-hand side of \eqref{eq:symplecticweight} equals the Hamming
weight of the vector $(\bm{a}|\bm{b})=\phi(X^{\bm{a}}Z^{\bm{b}})$ when
considered as a vector of length $n$ over $\F_q\times\F_q$, i.\,e.,
$\left((a_1|b_1),\ldots,(a_n|b_n)\right)$.

\subsection{Stabilizer Codes and Classical Codes}
With this preparation, we are ready to define the class of stabilizer
codes.
\begin{definition}[stabilizer code]
A stabilizer quantum error-correcting code
$\mathcal{C}=(\!(n,K,d)\!)_q$ is the common eigenspace (with
eigenvalue $+1$) of an Abelian subgroup $\mathcal{S}$ of the $n$-qudit
Pauli group $\mathcal{P}_n$ that does not contain a non-trivial
multiple of identity. The dimension of the code is
$K=\dim\mathcal{C}=q^n/|\mathcal{S}|$.
\end{definition}
Note that a stabilizer code is a complex vector space. In the
literature, the term ``stabilizer code'' is quite often used for the
corresponding classical additive code obtained via the homomorphism
$\phi$:
\begin{definition}[classical stabilizer code]
The image $C=\phi(\mathcal{S})$ of the stabilizer subgroup
$\mathcal{S}$ is an additive ($\F_p$-linear) code of length $n$ over
the alphabet $\F_q\times\F_q$ respectively an additive code of length
$2n$ over $\F_q$.
\end{definition}
As the stabilizer subgroup $\mathcal{S}$ is Abelian, we have
\begin{lemma}
The classical stabilizer code $C=\phi(\mathcal{S})$ is self-orthogonal
with respect to the symplectic form \eqref{eq:symplectic}.
\end{lemma}
There is a one-to-one correspondence between the Abelian subgroups of
$\mathcal{P}_n$ having trivial intersection with the center
$Z(\mathcal{P}_n)$ and the additive codes that are self-orthogonal
with respect to the symplectic form \eqref{eq:symplectic}.  Special
cases include $\F_q$-linear codes and $\F_{q^2}$-linear codes when
choosing a suitable basis of $\F_{q^2}$ over $F_q$ and identifying
$\F_{q^2}$ with $\F_q^2$.  For $\F_q$-linear codes, the codes are
self-orthogonal with respect to the symplectic form $(\bm{a}|\bm{b}) *
(\bm{a}'|\bm{b}') = \bm{a}'\cdot\bm{b}-\bm{a}\cdot\bm{b}'$, omitting
the trace. For $\F_{q^2}$-linear codes, the codes are self-orthogonal
with respect to the Hermitian form $\bm{x}*\bm{y}=\sum_{i=1}^n
x_iy_i^q$ for vectors $\bm{x},\bm{y}\in\F_{q^2}^n$. More details can
be found in \cite{AsKn01,KKKS06}.

The stabilizer subgroup $\mathcal{S}$ does not only define the
stabilizer code $\mathcal{C}$ itself, it also gives rise to an
orthogonal decomposition of the whole Hilbert space
$\left(\C^q\right)^{\otimes n}$ into spaces of equal dimension $K$
labeled by the characters of the group $\mathcal{S}$. Fixing a set of
$\kappa$ independent generators $s_i$ of the group $\mathcal{S}$ of
size $p^\kappa$, the characters correspond to the tuple of eigenvalues
$(\lambda_1,\ldots,\lambda_\kappa)$ of the generators.  As mentioned
above, one can find a measurement with the corresponding eigenspaces.
Although the operators $s_i$ are in general unitary and not
self-adjoint operators, one finds the terminology ``measuring the
stabilizer (generators) $s_i$'' in the literature for the corresponding
measurement.

In the following, we sketch how the correction of quantum errors is
related to the classical stabilizer code.  We consider an error
operator $E=X^{\bm{a'}}Z^{\bm{b'}}\in\mathcal{P}_n$. Furthermore, we
fix a generator\footnote{For simplicity, we assume that the phase
  factor $\omega_p^\gamma=1$.} $s_i=X^{\bm{a}}Z^{\bm{b}}$ of the
stabilizer group $\mathcal{S}$. For any pure state
$\ket{c}\in\mathcal{C}$ of the code, we compute
\begin{alignat}{5}
E\ket{c}
=Es_i\ket{c}
&{}=X^{\bm{a'}}Z^{\bm{b'}}X^{\bm{a}}Z^{\bm{b}}\ket{c}\nonumber\\
&{}=\omega_p^{(\bm{a}'|\bm{b'})*(\bm{a}|\bm{b})} X^{\bm{a}}Z^{\bm{b}}X^{\bm{a'}}Z^{\bm{b'}}\ket{c}
=\omega_p^{(\bm{a}'|\bm{b'})*(\bm{a}|\bm{b})} s_i E\ket{c}.
\end{alignat}
Hence the states $E\ket{c}$ are eigenvectors of the stabilizer
generator $s_i$ with eigenvalue
$\omega_p^{(\bm{a}'|\bm{b'})*(\bm{a}|\bm{b})}$. Therefore, an error
$E$ for which $\phi(E)*\phi(s_i)\ne 0$ for some $s_i\in\mathcal{S}$
can be detected, as the spaces $\mathcal{C}$ and $E\mathcal{C}$
correspond to different eigenvalues of $s_i$ and are hence orthogonal
to each other. On the other hand, undetectable errors that act
non-trivially on the code are exactly those with $\phi(E)*\phi(s)=0$
for all $s\in\mathcal{S}$, but $E\notin\mathcal{S}$.  The errors $E$ with
$\phi(E)*\phi(s)=0$ form a group, the centralizer of $\mathcal{S}$ in
$\mathcal{P}_n$. In this particular case, the centralizer and the
normalizer agree, and usually the term \emph{normalizer group}
$\mathcal{N}$ is used.

The homomorphism $\phi$ captures this situation as well:
\begin{definition}[classical normalizer code]
The image of the normalizer group $\mathcal{N}$ associated with a
stabilizer code with stabilizer group $\mathcal{S}$ under the
homomorphism $\phi$ is an additive code $C^*$ that is the dual code of
the classical stabilizer code $C=\phi(\mathcal{S})$ with respect to
the symplectic form \eqref{eq:symplectic}, i.\,e.,
\begin{alignat}{5}
C^*=\{ (\bm{a}'|\bm{b}')\colon
(\bm{a}'|\bm{b}')*(\bm{a}|\bm{b})=0\quad\text{for all $(\bm{a}|\bm{b})\in C$}\}.
\end{alignat}
\end{definition}
In summary, we can express the minimum distance of a stabilizer code
$\mathcal{C}$ in terms of the associated classical codes:
\begin{theorem}
The minimum distance $d$ of a stabilizer code
$\mathcal{C}=(\!(n,K,d)\!)_q$ with associated classical stabilizer code
$C=\phi(\mathcal{S})$ is given by 
\begin{alignat}{5}
d=\min\{\wgt(\bm{a}|\bm{b})\colon (\bm{a}|\bm{b})\in C^*\setminus C\}
\ge\min\{\wgt(\bm{a}|\bm{b})\colon (\bm{a}|\bm{b})\in C^*\}=d_{\text{min}}(C^*),
\end{alignat}
where we use the Hamming weight for codes over the alphabet
$\F_q\times \F_q$ (see \eqref{eq:symplecticweight}). A code with
$d=d_{\text{min}}(C^*)$ is called \emph{pure}, otherwise impure.
\end{theorem}

As outlined above, error correction is based on performing a
measurement that yields the tuple of eigenvalues
$(\lambda_1,\ldots,\lambda_\kappa)$ of a set of $\kappa$ independent generators
$s_i$ of the stabilizer group $\mathcal{S}$.
For an error $E$, the eigenvalues are 
\begin{alignat}{5}
\lambda_i=\omega_p^{\phi(E)*\phi(s_i)}.
\end{alignat}
The vectors $\phi(s_i)$ are generators of the additive code $C$ and
hence give rise to check equations for the dual code $C^*$. The task
is to determine the error $E$ using the classical syndrome vector
\begin{alignat}{5}
\left(\phi(E)*\phi(s_1),\ldots,\phi(E)*\phi(s_\kappa)\right)\in\F_p^\kappa,
\end{alignat}
which can be solved by a classical decoding algorithm for the code
$C^*$. Note that given $\phi(E)$, the error operator $E$ is only
determined up to a phase factor $\omega_p^\gamma$, but this is irrelevant.

Most stabilizer codes in the literature are constructed from codes
that are not just additive, but linear over $\F_q$ or $\F_{q^2}$.  In
this case, the size of the stabilizer group is an integral power of $q$.
Then the dimension $K$ of the stabilizer code $\mathcal{C}=(\!(n,K,d)\!)_q$ is
an integral power $K=q^k$ of $q$ as well, and the notation
$\mathcal{C}=[\![n,k,d]\!]_q$ is used. The parameter $k$ is referred
to as the number of encoded or logical qudits of the code.

\section{Secondary Constructions for Quantum Codes}
\subsection{Trivial Constructions}
First, we consider some trivial propagation rules for the parameters
of quantum error-correcting codes. For qubit codes, such rules have
been stated in \cite[Theorem 6]{CRSS98}. The first set of rules apply
to all QECCs.
\begin{theorem}
Assume that a quantum code $\mathcal{C}=(\!(n,K,d)\!)_q$ with $d>1$ exists. Then the
following QECCs exist as well:
\begin{enumerate}
\item $\mathcal{C}'=(\!(n,K',d)\!)_q$ for all
  $1<K'\le K$ (subcode)
\item $\mathcal{C}'=(\!(n',K,d)\!)_q$ for all $n'\ge n$ (lengthening)
\item $\mathcal{C}'=(\!(n-1,K,d-1)\!)_q$ (puncturing)
\end{enumerate}
\end{theorem}
\begin{proof}\ 
\begin{enumerate}
\item Any subspace $\mathcal{C}'$ of dimension $K'$ of the code
  $\mathcal{C}$ has clearly at least the same minimum distance.
\item In order to increase the length of a code, one simply takes the
  tensor product of any state $\ket{c}\in\mathcal{C}$ of the code with
  a fixed quantum state $\ket{\psi_0}$ with $n'-n$ qudits. When
  decoding the code $\mathcal{C}'$, one can just discard the last
  $n'-n$ qudits and decode the first $n$ qudits with respect to the
  original code $\mathcal{C}$. Then one takes the tensor product with
  the known fixed state $\ket{\psi_0}$. The resulting code
  $\mathcal{C}'$ will be impure, as there are error operators of
  weight less than $d$ that act only on the last $n'-n$ qudits and
  that stabilize the code.
\item Fix a pure state $\ket{\psi_0}$ of a single qudit and consider
  the projection operator $P_0=I_{q^{n-1}}\otimes
  \ket{\psi_0}\bra{\psi_0}$. As $\wgt(P_0)=1<d$,
  eq. \eqref{eq:KnillLaflamme2} implies that
  $\bra{c_i}P_0\ket{c_i}=\alpha$ for all basis states $\ket{c_i}$ of
  the code $\mathcal{C}$.  We can choose the state $\ket{\psi_0}$ such
  that $\alpha\ne 0$. After renormalization, the images of the basis
  states $\ket{c_i}$ under the projection $P_0$ have the form
  $\ket{c_i'}\otimes\ket{\psi_0}$. The new code $\mathcal{C}'$ is
  spanned by the states $\ket{c_i'}$ on $n-1$ qudits.

  Assume that an error $E'$ acts on the punctured code
  $\mathcal{C}'$. Considering the tensor product of the received state
  with the fixed state $\ket{\psi_0}$ is equivalent to the error
  $E=E'\otimes \ket{\psi_0}\bra{\psi_0}$ acting on the original code
  $\mathcal{C}$. For the original code, the Knill-Laflamme
  conditions hold for all errors with $\wgt(E)<d$. As $\wgt(E)=\wgt(E')+1$,
  for the punctured code $\mathcal{C}'$ the Knill-Laflamme conditions
  will hold for all errors with $\wgt(E')<d-1$, i.\,e., the minimum
  distance of the punctured code is at least $d-1$.
\end{enumerate}
\end{proof}
We can also trivially combine two codes.
\begin{theorem} Assume that quantum error-correcting codes
  $\mathcal{C}_1=(\!(n_1,K_1,d_1)\!)_q$ and
  $\mathcal{C}_2=(\!(n_2,K_2,d_2)\!)_q$ exist. Then a code
  $\mathcal{C}'=(\!(n_1+n_2,K_1K_2,\min\{d_1,d_2\})\!)_q$ exist as well.
\end{theorem}
\begin{proof}
The code $\mathcal{C}'$ is the tensor product of the codes
$\mathcal{C}_1$ and $\mathcal{C}_2$. The total number of qudits is
$n_2+n_2$, and the dimension of the tensor product is
$K_1K_2$. Decoding the two blocks independently implies that the
minimum distance of the tensor product is at least the minimum of
$d_1$ and $d_2$.
\end{proof}
Using the isomorphism between $\C^{q^m}$ and
$\left(\C^q\right)^{\otimes m}$, one can obtain codes for subsystems
of dimension $q$ from codes for subsystems of dimension $q^m$:
\begin{theorem} Assume that a quantum error-correcting code
  $\mathcal{C}=(\!(n,K,d)\!)_{q^m}$  exists. Then a code
  $\mathcal{C}'=(\!(mn,K,d'\ge d)\!)_q$ exist as well.
\end{theorem}
\begin{proof}
The code $\mathcal{C}'$ is the same complex vectors space as the
original code $\mathcal{C}$, but considered as a subspace of
$\left(\C^q\right)^{\otimes mn}$. The weight of an error does not
decrease under this expansion, and hence the minimum distance $d'$ of the
expanded code is not smaller than the original minimum distance $d$.
\end{proof}

\subsection{Shortening the Classical Stabilizer Code}
The list of trivial propagation rules does not include the analog of
shortening of classical block codes which reads
$C=(n,M,d)_q\Rightarrow C'=(n-1,M/q,d)_q$. The shortened code $C'$ is
obtained by first taking all codewords of $C$ with a fixed symbol at
the last position (usually the symbol $0$) and then discarding that
symbol.  When we apply this operation to the classical stabilizer
code, we obtain the following.
\begin{theorem}
Assume that a pure stabilizer code $\mathcal{C}=(\!(n,K,d)\!)_q$ with
$d>1$ and $q=p^m$ exists. Then a stabilizer code
$\mathcal{C}'=(\!(n-1,qK,d-1)\!)_q$ exists as well.
\end{theorem}
\begin{proof}
Let $C=\phi(\mathcal{S})=(n,p^\kappa,\delta)_q$ be the classical
stabilizer code associated with $\mathcal{C}$.  We have
$K=q^n/p^\kappa$.  Shortening the code $C$ yields an additive code
$C'=(n-1,p^{\kappa'},\delta)_q$ which remains self-orthogonal with
respect to the symplectic form.  The size of the shortened code
depends on the number of different symbols at the shortened position.
As the minimum distance of the symplectic dual code $C^*$ equals
$d>1$, all $q^2$ pairs $(a,b)\in\F_q\times\F_q$ occur at every
position $C$. Assuming the contrary, the code $C_1$ of length $1$
obtained by puncturing the code $C$ at all but one position has size
$p^\mu$ with $\mu< 2m$. Then the symplectic dual $C_1^*$ of size
$2m-\mu$ contains a non-zero codeword. The code $C_1^*$ equals the
code obtained by shortening $C^*$ at all but the corresponding
position. Hence, the minimum distance of $C^*$ would be $1$, a
contradiction. Therefore, the size of the code shortened code $C'$ is
$p^{\kappa-2m}$.  The minimum distance of $(C')^*$ is at least
$d-1$. For the dimension $K'$ of the new stabilizer code
$\mathcal{C}'$ we compute
\begin{alignat}{5}
K'=\frac{q^{n-1}}{p^{\kappa-2m}}=\frac{p^{2m}}{q}\frac{q^n}{p^\kappa}=qK.
\end{alignat}
\end{proof}
While we have stated the result only for stabilizer codes, this
propagation rule can be applied to all quantum codes \cite{HuGr20}.
For $\F_q$-linear codes, the propagation rule reads
$[\![n,k,d]\!]_q\Rightarrow [\![n-1,k+1,d-1]\!]_q$. So the quantity
$n+k$ is preserved. The minimum distance is decreased by one while the
number of logical qudits is increased by one when reducing the length
by one.

\subsection{Shortening Stabilizer Codes}\label{sec:puncturecode}
In certain cases, a propagation rule similar to that of shortening
classical codes is possible. For stabilizer codes, one would shorten
the classical normalizer code, reducing the length while preserving
the minimum distance. Shortening of the classical normalizer code
corresponds to puncturing of the classical stabilizer code. In
general, the punctured code is no longer self-orthogonal. The idea is
to multiply the coordinates of the punctured code by suitably chosen
scalars to make the code self-orthogonal. The construction was stated
for prime dimensions $q=p$ in \cite{Rai99:nonbinary}.
\begin{definition}[Rains puncture code]
Let $C=(n,p^\kappa)_q$ be an additive code over $(\F_q\times \F_q)^n$.
The puncture code $P(C)$ is the Euclidean dual of the additive code
over $\F_p$ obtained by evaluating the symplectic form
\eqref{eq:symplectic} on each coordinate of $C$ for all pairs of codewords, i.\,e.,
\begin{alignat}{5}
 P(C)=\left\langle \left(\tr(a_1 b'_1-a_1'b_1),\ldots,\tr(a_n b'_n-a_n'b_n)\right)
   \colon (\bm{a}|\bm{b}),(\bm{a}'|\bm{b}')\in C\right\rangle^\bot.\label{eq:PC}
\end{alignat}
\end{definition}
\begin{theorem}\label{thm:shortening}
Assume that the puncture code $P(C)$ of an additive code
$C=(n,p^\kappa)_q$ contains a codeword of weight $r$. Then one can
derive an additive self-orthogonal code $C'=(r,p^{\kappa'})_q$ with
$\kappa'\le\kappa$ from $C$. The minimum distance of $(C')^*$ is not
smaller than the minimum distance of $C^*$.
\end{theorem}
\begin{proof}
Let $\bm{c}$ be a codeword of $P(C)\le\F_p^n$ of weight $r$. Without
loss of generality assume that the first $r$ coordinates of $\bm{c}$
are non-zero. Consider the code $\widetilde{C}$ obtained by
multiplying the first part of each pair of coordinate symbols
$(a_i|b_i)$ by $c_i$, i.\,e.,
\begin{alignat}{5}
\widetilde{C}=\left\{\left( c_1 a_1,\ldots,c_n a_n|b_1,\ldots,b_n\right)
   \colon (\bm{a}|\bm{b})\in C\right\}.
\end{alignat}
The code $C'$ is obtained by puncturing $\widetilde{C}$ at the last
$n-r$ positions. Using the definition \eqref{eq:PC} of the puncture
code and the $\F_p$-linearity of the trace, it follows that both
$\widetilde{C}$ and $C'$ are self-orthogonal. The size of $C'$ is not
larger than the size of the original code $C$.  Puncturing the code
$C$ corresponds to shortening the symplectic dual code, which implies
the lower bound on the minimum distance.
\end{proof}
\begin{corollary}
Assume that the puncture code of a stabilizer code
$\mathcal{C}=(\!(n,K,d)\!)_q$ contains a word of weight $r=n-s$. Then
a stabilizer code $\mathcal{C}'=(\!(n-s,q^{-s}K,d)\!)_q$ exists as well.
\end{corollary}
\begin{proof}
Using Theorem \ref{thm:shortening}, we only have to compute the
dimension of the stabilizer code derided from $C'$.  As $|C'|\le
p^\kappa=q^n/K$, we calculate
\begin{alignat}{5}
K'=\frac{q^{n-s}}{|C'|}\ge K q^{-s}.
\end{alignat}
\end{proof}
The remaining problem in algebraic coding theory is to determine which
weights occur in the puncture code.

\begin{remark}\label{rem:PC} As it can be seen from the proof of Theorem
  \ref{thm:shortening}, the definition of the puncture code can be
  generalized as follows:
\begin{alignat}{5}
 \widetilde{P}(C)=\bigl\{(c_1,\ldots,c_n)\in\F_q^n\colon
   \sum_{i=1}^n \tr\left(c_i(a_i b'_i-a_i'b_1)\right))=0\quad\text{for
     all $(\bm{a}|\bm{b}),(\bm{a}'|\bm{b}')\in C$}\bigr\}.
\end{alignat}
In comparison, the original definition \eqref{eq:PC} can be expressed as
\begin{alignat}{5}
 P(C)=\bigl\{(c_1,\ldots,c_n)\in\F_p^n\colon
   \sum_{i=1}^n c_i\tr\left((a_i b'_i-a_i'b_i)\right))=0\quad\text{for
     all $(\bm{a}|\bm{b}),(\bm{a}'|\bm{b}')\in C$}\bigr\}.
\end{alignat}
This generalization has yet to be investigated.
\end{remark}

\subsection{Extending Additive Codes}
Another possibility to obtain self-orthogonal additive codes is by
extension, i.\,e., increasing the length of a code $C=(n,p^\kappa)_q$ by
adding new symbols such that the resulting code $C'=(n',p^\kappa)_q$
becomes self-orthogonal.  This approach has been used in \cite{WRLS20}
to obtain stabilizer codes with improved parameters. The main problem
is, however, to control the minimum distance of the symplectic dual
$(C')^*$ of the extended code and hence the minimum distance of the
resulting stabilizer code.

We have found a new code $\mathcal{C}=[\![96,50,10]\!]_2$. The classical
normalizer code is an $\F_4$-linear code $C^*=[96,73,10]_4$ containing
its Hermitian dual. The code $C^*$ is obtained by a three-step variant
of Construction X (see, e.\,g., \cite{MS77}) using a cyclic code
$C_1=[93,73,8]_4$ with generator polynomial 
\begin{alignat}{5}
g_0(x)={}&(x^5 + x^2 + \alpha)
(x^5 + x^3 + x^2 + x + 1)\nonumber\\
&(x^5 + \alpha x^4 + x^2 + \alpha x + \alpha^2)
(x^5 + \alpha x^4 + \alpha^2x^3 + x^2 + \alpha^2),
\end{alignat}
where $\alpha$ is a generator of $\F_4$, 
and subcodes with generator polynomial
$g(x)(x-1)^a(x-\alpha)^b(x-\alpha^2)^c$, $a,b,c\in\{0,1\}$. It should
be noted that shortening the stabilizer of the new code
$\mathcal{C}=[\![96,50,10]\!]_2$ three times yields a
code $\mathcal{C}'=[\![93,53,8]\!]_2$ whose classical normalizer code is
the code $C_1$.

\subsection{Quantum Construction X}\label{sec:QConstructionX}
Lisonek and Singh \cite{LiSi14} proposed a construction of Hermitian
self-orthogonal $\F_{q^2}$-linear codes applying Construction X to an
arbitrary linear code of rate at most $1/2$ over $\F_{q^2}$ using a
trivial code of distance one as auxiliary code.  While Lisonek and
Singh focused on the case $q=2$, the construction has natural
extensions to linear codes over $\F_{p^2}$ \cite{DGG15} as well as
quadratic extension fields $\F_{q^2}$ in general \cite{ELOS19}. In
\cite{LiDa18} the construction was extended to additive codes over
$\F_4$.  We present the main result in the version of \cite[Theorem
  2]{ELOS19}:
\begin{theorem}
For a linear code $C=[n,k]_{q^2}$, let $e:=k-\dim_{\F_{q^2}}(C\cap
C^*)$. Then there exists a stabilizer code
$\mathcal{C}=[\![n+e,n-2k+e,d]\!]_q$ with $d\ge
\min\{d(C^*),d(C+C^*)+1\}$.
\end{theorem}
\begin{proof} We sketch the main idea. The code $C\cap C^*$ is a
  linear subcode of $C$ of co-dimension $e$ that is Hermitian
  self-orthogonal. One can show that it is possible to find a basis of
  the complement $\tilde{C}$ of $C\cap C^*$ in $C$ such that
  Construction X applied to $C$ and $C\cap C^*$ using a trivial
  auxiliary code $C_{\text{aux}}=[e,e,1]_{q^2}$ yields a linear
  Hermitian self-orthogonal code $C'=[n+e,k]_{q^2}$. The co-dimension
  of $C^*$ in $(C\cap C^*)^*=C+C^*$ is $e$ as well. It turns out that
  the Hermitian dual code $(C')^*$ is spanned by the codewords of $C$
  with $e$ zeros appended, as well as the vectors corresponding to the
  juxtaposition of $\widetilde{C}$ and $C_{\text{aux}}$. This implies
  the lower bound on the minimum distance.
\end{proof}
Note that the linear code $C\cap C^*$ of dimension $k-e$ in the
proof yields a stabilizer code $\mathcal{C}_0=[\![n,n-2k+2e,d_0]\!]_q$
where $d_0\ge d(C+C^*)$. For $e=1$, the parameters of the code
$\mathcal{C}_0$ are exactly those which one obtains by shortening the
classical stabilizer code of the code $\mathcal{C}$ obtained by
quantum Construction X. Hence, one can consider quantum Construction X
in some sense as the inverse of shortening the classical stabilizer
code.

Starting with a quantum code $\mathcal{C}=[\![n,k,d]\!]_q$, one
considers the classical normalizer code $C^*$.  Then one seeks a
subcode $C_1^*$ of $C^*$ of small co-dimension, but with a minimum distance
that is at least $d+1$.  In general, the code $C_1^*$ will not contain
its symplectic dual $C_1$. Applying quantum Construction X to $C_1$,
one will obtain a quantum code $\mathcal{C}'=[\![n+e,k-e,d+1]\!]_q$ for
some $e$.

\section{Conclusions}
Starting with the characterization of general quantum error-correcting
codes, we have discussed how algebraic coding theory fits into that
picture via the theory of stabilizer codes.  Classical codes that are
self-orthogonal with respect to a symplectic form correspond to
Abelian subgroups of the generalized Pauli group, defining quantum
codes via their joint eigenspaces.  At the same time, using a local
error model for quantum codes naturally corresponds to the Hamming
metric for classical codes.  By now, there are numerous publications
from the algebraic coding theory community on the construction
self-orthogonal classical codes that give rise to good quantum codes.
There have also been attempts to adopt specific concepts from
algebraic coding theory---like codes for different metrics than the
Hamming metric---to quantum codes, missing to relate the concepts to
quantum mechanics.  On the other hand, we have only considered the
depolarizing channel and the quantum erasure channel here, together
with their $n$-fold uses. For other quantum channels, one has to
investigate what their correspondence in algebraic coding theory is
and how one can find good codes for that situation.

Complementing the direct construction of classical self-orthogonal
codes leading to stabilizer codes with good parameters, we presented
the notion of the puncture code in Section \ref{sec:puncturecode} and
quantum Construction X in Section \ref{sec:QConstructionX}.  The
puncture code has been used to show the existence of many quantum MDS
codes \cite{GrRo15}.  In general it is an open question how one can
determine the weights in the puncture code for certain classes of
codes.  The generalization mentioned in Remark \ref{rem:PC} is worth
to be further explored.

For quantum Construction X, the main question is how to find nested
classical codes that result in quantum codes with good parameters.
Another promising concept that we have not discussed here is
generalized concatenation for quantum codes \cite{GSZ09}.
Concatenation can also be a method to design good codes for specific
quantum channels.

\section*{Acknowledgments}
The author acknowledges discussions with Frederic Ezerman, Petr
Lison\v{e}k, Buket {\"O}zkaya, and Martin R\"otteler, as well as
discussions during the Oberwolfach Workshop 1912 on Contemporary
Coding Theory, March 2019.  The `International Centre for Theory of
Quantum Technologies' project (contract no. 2018/MAB/5) is carried out
within the International Research Agendas Programme of the Foundation
for Polish Science co-financed by the European Union from the funds of
the Smart Growth Operational Programme, axis IV: Increasing the
research potential (Measure 4.3).


\end{document}